\newcommand{\be}{\begin{equation}}
\newcommand{\ee}{\end{equation}}
\newcommand{\bea}{\begin{eqnarray}}
\newcommand{\eea}{\end{eqnarray}}
\newcommand{\ba}[1]{\begin{array}{#1}}
\newcommand{\ea}{\end{array}}

\documentclass[pra,aps,showpacs,twocolumn]{revtex4-1}
\usepackage{times}
\usepackage{amssymb}
\usepackage{amsmath}
\usepackage{mathrsfs}
\usepackage{graphicx}
\usepackage{epsfig}
\usepackage{dcolumn}
\usepackage{color}
\usepackage{bm}

\begin{document}

\title{Photon Scattering from a System of Multi-Level Quantum Emitters. I. Formalism}
\author{Sumanta Das$^{1}$, Vincent E. Elfving$^{1}$, Florentin Reiter$^{2}$, and Anders S. S\o rensen$^{1}$}
\affiliation{$^{1}$Niels Bohr Institute, University of Copenhagen, Blegdamsvej 17, 2100 Copenhagen \O, Denmark\\
$^{2}$ Department of Physics, Harvard University, Cambridge, MA 02138, USA}
\date{\today}
\begin{abstract}
We introduce a formalism to solve the problem of photon scattering from a system of multi-level quantum emitters. Our approach provides a direct solution of the scattering dynamics. As such the formalism gives the scattered fields amplitudes in the limit of a weak incident intensity. Our formalism is equipped to treat both multi-emitter and multi-level emitter systems, and is applicable to a plethora of photon scattering problems including conditional state preparation by photo-detection. In this paper, we develop the general formalism for an arbitrary geometry. In the following paper (part II), we reduce the general photon scattering formalism to a form that is applicable to $1$-dimensional waveguides, and  show its applicability by considering explicit examples with various emitter configurations.
\end{abstract}

\maketitle

\section{Introduction}
Interaction of an electromagnetic field with quantum emitters is a subject of fundamental importance and forms the basis of quantum optics \cite{Agarwalbook}. During the past decades several fascinating phenomena like single photon superradiance, electromagnetically induced transparency, Anderson localization of light, Rydberg blockade and photon-photon interaction have been realized owing to such light-matter interactions \cite{ScullyPRL, PetruPRL, FleisRMP, MorNp, Molmer_rmp, Chang_rev}. With the advent of quantum information sciences, the investigation of light-matter coupling has received paramount interest. Efficient coupling of a single photon to a quantum emitter, is of central importance for future quantum technologies \cite{Kim08,Tey08, Brien09, Chang07, Hwang09}. The key challenge in achieving this is two-fold: on one hand we currently lack stable single photon sources while on the other hand, the probability that a single photon in a light beam interacts with a quantum emitter-like atom is very small \cite{Chang_rev}. 

Various physical systems ranging from atoms to nitrogen-vacancy centers in diamond and superconductors are been actively investigated to achieve strong and efficient light-matter coupling in the quantum regime \cite{Molmer_rmp, Haroche_rmp, Rempe_rmp, Peter_rmp, Anders_rmp, Kurt20, Brouri20, Yuan05, Fu08, Rebic09, Babi10, Liew10,  Bamba11, Maju12, Pey12, Loo13, Baur14, Tiecke14, Giesz16}. These investigations can be broadly classified into two approaches, one concerns ensembles of quantum emitters to collectively increase the cross-section of light-matter interaction, while the other involves tight confinement of the electromagnetic field in cavities or other dielectric media like superconducting transmission lines, nanowires and waveguides containing the quantum emitters. To be able to harness such interfaces, a thorough understanding of the dynamics of light scattering from (multiple) quantum emitters in such dielectric media is required. The importance of this problem is acknowledged by the extensive investigations done on this topic over the span of the last decade \cite{Shen05, Shields07, Houck07, Yud08, With10, Zheng11}. These works typically restricts themselves either to the case of $1$-dimensional dielectric medium or consider the simplest case of scattering from a single or multiple two level emitters \cite{Shen05, Houck07, Yud08, With10, Zheng11, Zhou08, Lalu13, Roy11, Laakso14}. Thus, the problem of photon scattering from multiple multilevel quantum emitters in a general dielectric medium remains unsolved in general. 
                                              
Motivated by this, in this article we introduce a photon scattering relation for a weak field scattered off a generic system of multi-level emitters embedded in an general $3$-dimensional dielectric medium within the Markov approximation. We develop our formalism from first principles in the Heisenberg picture and obtain the scattering relation between the input and output electromagnetic field operators in terms of the inverse of a non-Hermitian Hamiltonian of the system and the emitters collective ground state coherence. Most importantly, our formalism can deal with any possible complex intra- and inter-emitter dynamics, as long as the non-Hermitian Hamiltonian can be inverted. In spirit, our photon scattering relation has similarities to the well known input-output formalism of quantum optics that is extensively used in cavity QED \cite{Walls}. The input-output formalism, however, only gives the dynamical equations of motion, which still needs to be solved. On the contrary, our approach provides a full solution to the scattering problem. The resulting formalism may thus also be used to provide a solution to the scattering dynamics within the context of cavity QED. 

To solve for the dynamics of the emitters and thereby the response of the scattering medium on the incident photons, we make the following key assumption. The system dynamics can be divided into two different time scales: $1)$ a fast dynamics within an excited state subspace, e.g. large detunings or fast decay rates, and $2)$ a slower time scale associated with the excitation out of the ground subspace, e.g., due to a low rate of incoming photons. In this situation we can eliminate the excited state and obtain an effective ground state dynamics as well as the full scattering relation. This procedure is similar to the well-known technique of adiabatic elimination and can also be seen as the generalization of the effective operator (EO) technique presented in Ref. \cite{Reiter12} to quantum fields. Note that this procedure is only applicable in the regime where the emitters have a negligible probability to be doubly-excited. We satisfy this condition in our approach by restricting the scattering formalism to the weak excitation regime \cite{Deutsch95, Chang12}. As such, our formalism cannot account for fast non-linear processes involving multi-photon scattering arising from strong input intensities \cite{Fan10, Can15, Shi15}. On the other hand, the non-linear effect associated with the slow evolution of the ground state coherence, e.g., through Raman transitions \cite{Eduarxiv17}, is completely accounted for by EO equations of motion. The EO's include all induced properties like phases, decay, coupling, detuning, shifts of energy level etc. via the elements of a non-Hermitian Hamiltonian defined for the system. The method thus allows for the description of scattering involving a number of interesting photon states including the single-photon and weak coherent states. 

The article is organized as follows: In Sec. II we discuss the system and introduce the Hamiltonian. In Sec III we then derive the equation of motions for the multi-emitter system in the Heisenberg picture. In Sec. IV we find the solution of the single excitation coherence in terms of a non-Hermitian Hamiltonian of the system. In Sec V we use the effective operator method to eliminate the excited state manifold and develop a photon scattering relation in terms of the input field and the ground state dynamics of the emitters. In Sec. VI we derive the effective operator master equation for evaluating the system's ground state dynamics. Finally, in Sec. VII we  summarize our findings. In a subsequent paper we use the general photon scattering relation introduced here to derive a similar relation among the input and output field for $1$-dimensional dielectric medium (e.g. waveguides). Furthermore, we consider several explicit examples with various emitter configurations to show how to apply the photon scattering relation to find reflection and transmission amplitudes of the scattered photon. Readers specifically interested in the application of the formalism are encouraged to consult the second part of this series.
\section{System Hamiltonian}
\begin{figure}
\label{figure1}
\includegraphics[height = 7 cm]{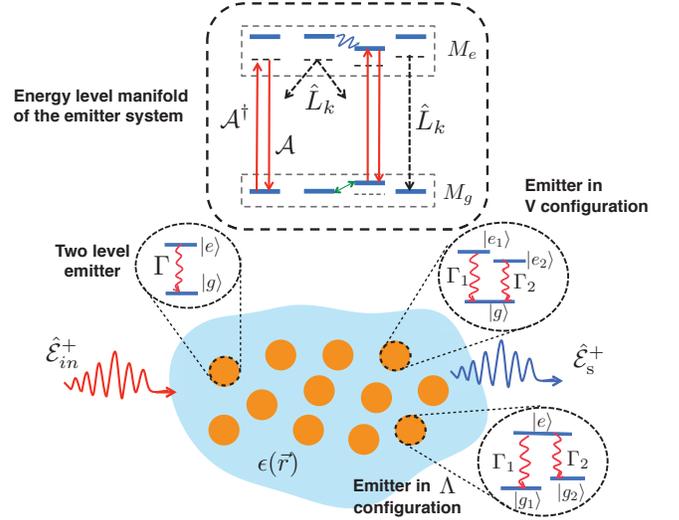}
\caption{Schematic of photon scattering from a generic system of emitters distributed in some dielectric medium with a spatial dependent electric permittivity $\epsilon(\vec{r})$. The emitters can be either a simple two level system with a decay rate $\Gamma$ or have multiple levels. The emitters are assumed to consists of two separated subspaces, an excited state manifold $M_e$ with excited states $|e_{m}\rangle ~(m = 1,2,.....,n)$ and corresponding decay rates $\Gamma_{m}$, and a ground state manifold $M_g$ with ground states $|g_{m}\rangle$. We assume the couplings between the two manifolds $V_{\pm}$ to be perturbative and model the excited states decay by Lindblad operators $\hat{L}_{k}$. The couplings within the excited and ground manifold are shown by the wiggling and straight arrow-headed lines, respectively. The mode operator for an incident field is represented by $\hat{\mathcal{E}}^{+}_{in}$, while the scattered outgoing field is given by the mode operator $\hat{\mathcal{E}}^{+}_{s}$. }
\end{figure}
In this section we introduce the model and the Hamiltonian that will be used to investigate the scattering of photons interacting with the quantum emitters. We consider a generic system of quantum emitters $j$ each located at the positions $\vec{\text{r}}_{j}$ in some dielectric medium of spatially dependent electric permittivity $\epsilon(\vec{r})$, as shown schematically in Fig. 1. We assume that the Hilbert space spanned by the states of the emitters can be separated into two subspaces, an excited subspace $M_e$ and a ground subspace $M_g$ formed by a manifold of excited states $\{|e\rangle\}$ and ground states $\{|g\rangle\}$, respectively. Hence the $M_e$ subspace comprises of the various possible combination of the emitter states with at least one emitter in the excited state, while $M_g$ comprises of the various possible combinations of the ground states of the emitters. As an example, consider the case of the emitter system comprising a two-level emitter labeled A and located at $\vec{\text{r}}_{\text{A}}$, and a three-level emitter in the $\Lambda$ configuration  labeled B and located at $\vec{\text{r}}_{\text{B}}$. The excited-state subspace $M_e$ is then formed by the manifold of four states given by $\{|e_\text{A},e_\text{B}\rangle, |e_\text{A},g_{1,\text{B}}\rangle, |e_\text{A},g_{2,\text{B}}\rangle, |g_{\text{A}},e_\text{B}\rangle\}$, while the ground-state subspace $M_g$ is formed by a manifold of two states given by $\{|g_\text{A}, g_{1,\text{A}}\rangle, |g_\text{A}, g_{2,\text{B}}\rangle\}$. In general, the manifold $\{|e\rangle\}$ can comprise of all possible excited states, $\{\otimes^{N}_{j=1}|e_{l,j}\rangle,\otimes^{N-1}_{j=1}|e_{l,j}\rangle|g_{m,N}\rangle,..., |e_{l,N}\rangle\otimes^{N-1}_{j = 1}|g_{m,j}\rangle\}$. Here, $N$ is the number of emitters, while the indices, $(l,m)$ in the subscripts of the excited $|e\rangle$ and ground $|g\rangle$ state correspond to the different energy levels within individual emitters. Note that $\{|e\rangle\}$ in addition can also be spanned by collective states of the form $\frac{1}{\sqrt{N}}\left(|e_{l,1}\rangle\otimes^{N}_{j = 2}|g_{m,j}\rangle+...+|e_{l,N}\rangle\otimes^{N-1}_{j = 1}|g_{m,j}\rangle\right)$. The ground state manifold $\{|g\rangle\}$ on the other hand corresponds to states of the form $\otimes^{N}_{j}|g_{m,j}\rangle$.  We emphasize that due to the generality of our model, the results of the current work can be applied to a plethora of quantum emitters like atoms, molecule, quantum dots, superconducting qubit, and nitrogen vacancies. 

We next consider the interaction of the emitters in the dielectric medium with an incoming light field represented by the $3$D electric field $\vec{E}(\vec{r},t)$. The Hamiltonian of our model system is then given by $\mathcal{H} = \mathcal{H}_{F}+\mathcal{H}_{c}+\mathcal{H}_{I}$, where the free-field Hamiltonian of the multimode electromagnetic field is given by $\mathcal{H}_{F} = \sum_{k} \hbar\omega_{k}\left (a^\dagger_k a_k+\frac{1}{2}\right)$ and $\mathcal{H}_{c}$ is the Hamiltonian of the emitters including intra- and inter-emitter interactions that are not mediated by the light. Here $a_{k} (a^\dagger_{k})$ is the bosonic field mode annihilation (creation) operator of the $k^{\text{th}}$ mode with frequency $\omega_{k}$. The total interaction Hamiltonian for our model system is given by $\mathcal{H}_{I} = \sum_{j}\mathcal{H}^{j}_{I}$, where $\mathcal{H}^{j}_{I}$ is the interaction Hamiltonian describing the coupling of the $3$D multimode electromagnetic field with an emitter located at a spatial position $\vec{r}_{j}$. Note that in the total Hamiltonian $\mathcal{H}$ only the interaction Hamiltonian $\mathcal{H}_{I}$ leads to coupling between the ground and excited subspaces $M_g$ and $M_e$.

In the rotating wave approximation, $\mathcal{H}^{j}_{I}$ can be written as
\bea
\label{eq1}
\mathcal{H}^{j}_{I} & = &-\vec{d}^{j}\cdot\vec{E}(\vec{r}_{j},t)\nonumber\\
&=&-i\sum_{k}\sum_{e,g}\sqrt{\frac{\hbar\omega_k}{2}}[\vec{d}^{j}_{eg}\cdot\vec{F}_{k}(\vec{r}_j)]\hat{\sigma}_{ge}\hat{a}_{k}+H.c.,
\eea
where $\vec{d}^{j}$ is the $j^{\text{th}}$ emitter's dipole operator defined by $\vec{d}^{j} = \sum_{eg}(\vec{d}^{j}_{eg}|e\rangle\langle g|+\vec{d}^{j}_{ge}|g\rangle\langle e|)$ with $\vec{d}^{j}_{eg} = \langle e|\vec{d}^{j}|g\rangle$ being the dipole moment of the transition $|e\rangle\leftrightarrow|g\rangle$, between the excited $|e\rangle$ and ground states $|g\rangle$, that were introduced in the previous paragraph. 
Note that since $|e\rangle$ and $|g\rangle$ can be collective states of the total system, there can be multiple dipole transitions between the states $|e\rangle$ and $|g\rangle$. We define our emitter-raising (lowering) operator as  $\hat{\sigma}_{ge} = |e\rangle \langle g|  (\hat{\sigma}_{eg} = |g\rangle\langle e|)$ between the excited and ground states. Note that we follow a non-standard definition of these operators with the bras and kets of the operators in opposite order.  We chose this convention to ensure that, e.g., element of the density matrix $\rho_{eg}$ can be found as the expectation value $\langle \hat{\sigma}_{eg}\rangle$. This convention will simplify the expressions below. The operators then satisfy the angular momentum commutation relation $[\hat{\sigma}_{eg},\hat{\sigma}_{g'e'}] = -2\sigma_{z}\delta_{ee'}\delta_{gg'}$, where $\sigma^{z} = \left(|e\rangle\langle e|-|g\rangle\langle g|\right)$. 

The $3$-dimensional quantized multimode electric field has the form $\vec{E}(\vec{r},t)  = \hat{\vec{\mathcal{E}}}^{+}(\vec{r},t)+\hat{\vec{\mathcal{E}}}^{-}(\vec{r},t)$ \cite{Scullyb} with
\bea
\label{eq2}
\hat{\vec{\mathcal{E}}}^{+}(\vec{r},t) & = &i\sum_{k}\sqrt{\frac{\hbar\omega_k}{2}}\vec{F}_k(\vec{r})\hat{a}_k(t),
\eea 
and $\hat{\vec{\mathcal{E}}}^{-}(\vec{r},t) = \left[\hat{\vec{\mathcal{E}}}^{+}(\vec{r},t)\right]^\dagger$. Here $\epsilon$ is the electric permittivity of the di-electric medium while $\vec{F}_k(\vec{r})$ is some general eigen-mode function corresponding to the $k^\text{th}$ mode of the electromagnetic field, satisfying the $3$-dimensional Maxwell's wave equation and the orthogonality relation \cite{Glauber}
\bea
\label{eq3}
\int~d\vec{r}\epsilon(\vec{r})\vec{F}_k(\vec{r})\vec{F}^\ast_{k'}(\vec{r}) = \delta_{kk'}.
\eea
The interaction Hamiltonian in Eq. (\ref{eq1}) is completely general and includes the multi-level (summation over $e, g$) structure of the quantum emitters as well as the multi-mode $3$D characteristic of the field. 

Our main interest here is the description of dissipative processes induced by the interaction with light. For this purpose the above Hamiltonian in the rotating wave approximation is sufficient. In addition to dissipation   the theory developed below will also include light induced dipole-dipole interactions among the emitters. For emitters separated by less than a wavelength the Hamiltonian in the rotating-wave approximation does not give the correct dipole-dipole interaction. In this case the interactions terms derived below should therefore be replaced by the appropriate expressions taking into account the full interaction \cite{Lehm70, Wubs04, Das08, Svid10, Miro13, Ott13}. Except for the light-induced dipole-dipole interaction we will not explicitly include any other direct interactions (like dipolar, Rydberg etc) between the emitters. However, such interactions can be included in $\mathcal{H}_c$. 
\section{The equations of motion}
We now investigate the dynamics of the multi-level emitters using the Hamiltonian introduced above in the Heisenberg picture. Using the total Hamiltonian $\mathcal{H}$ of the emitter-waveguide system we write the Heisenberg equations of motion for the field modes $\hat{a}_{k}$ and any general operator $\hat{\mathcal{O}}$ for the  emitters as 
\bea
\label{eq4}
\dot{\hat{a}}_k & = & -i\omega_k\hat{a}_{k} + \sum_{j,e,g}\sqrt{\frac{\omega_k}{2\hbar}}\left[\vec{F}^\ast_{k}(\vec{r}_j)\cdot\vec{d}^{j}_{ge}\right]\hat{\sigma}_{eg},
\eea
\bea
\label{eq5}
\dot{\hat{\mathcal{O}}} & = & \frac{i}{\hbar}\left[\hat{\mathcal{H}}_{c}, \hat{\mathcal{O}}\right] +\sum_{k}\sum_{j,e,g}\sqrt{\frac{\omega_k}{2\hbar}}\bigg(\bigg\{\vec{d}^{j}_{eg}\cdot\vec{F}_{k}(\vec{r}_j)\bigg\}\nonumber\\
&\times&\left[\hat{\sigma}_{ge},\hat{\mathcal{O}}\right]\hat{a}_{k}-\left\{\vec{F}^{\ast}_{k}(\vec{r}_j)\cdot\vec{d}^{j}_{ge}\right\}\hat{a}^\dagger_{k}\left[\hat{\sigma}_{eg},\hat{\mathcal{O}}\right]\bigg).
\eea
We formally integrate Eq. (\ref{eq4}) to get 
\bea
\label{eq6}
\hat{a}_k(t) & = &\hat{a}_{k}(0)e^{-i\omega_k t} +\sum_{j,e,g}\sqrt{\frac{\omega_k}{2\hbar}}\left[\vec{F}^\ast_{k}(\vec{r}_j)\cdot\vec{d}^{j}_{ge}\right]\nonumber\\
&\times&\int^{t}_{-\infty}e^{-i\omega_k(t-t')}\hat{\sigma}_{eg}(t')dt'. 
\eea
Substituting Eq. (\ref{eq6}) and its Hermitian conjugate into Eq. (\ref{eq5}) we then get
\begin{align}
\label{eq7}
{}&\dot{\hat{\mathcal{O}}}(t) = \frac{i}{\hbar}[\hat{\mathcal{H}}_{c}, \hat{\mathcal{O}}(t)] +\sum_{k}\sum_{j,e,g}\sqrt{\frac{\omega_k}{2\hbar}}\left\{\vec{d}^{j}_{eg}\cdot\vec{F}_{k}(\vec{r}_j)\right\}\nonumber\\
{}&\times[\hat{\sigma}_{ge}(t),\hat{\mathcal{O}}(t)]\hat{a}_{k_{F}}+\sum_{k,j,e,g}\left(\frac{\omega_k}{2\hbar}\right)\sum_{j',e',g'}\left\{\vec{d}^{j}_{eg}\cdot\vec{F}_{k}(\vec{r}_j)\right\}\nonumber\\
{}&\int^{t}_{-\infty}dt'[\hat{\sigma}_{ge}(t),\hat{\mathcal{O}}(t)]\left\{\vec{F}^{\ast}_{k}(\vec{r}_{j'})\cdot\vec{d}^{j'}_{g'e'}\right\}\hat{\sigma}_{e'g'}(t')e^{-i\omega_{k}(t-t')}\nonumber\\
{}&-\sum_{k}\sum_{j,e,g}\sqrt{\frac{\omega_k}{2\hbar}}\left\{\vec{F}^\ast_{k}(\vec{r}_j)\cdot\vec{d}^j_{ge}\right\}
\hat{a}^{\dagger}_{k_{F}}[\hat{\sigma}_{eg}(t),\hat{\mathcal{O}}(t)]\nonumber\\
{}&-\sum_{k,j,e,g}\left(\frac{\omega_k}{2\hbar}\right)\sum_{j',e',g'}\left\{\vec{d}^{j'}_{e'g'}\cdot\vec{F}_{k}(\vec{r}_{j'})\right\}\nonumber\\
{}&\int^{t}_{-\infty}dt'\hat{\sigma}_{g'e'}(t')\left\{\vec{F}^\ast_{k}(\vec{r}_j)\cdot\vec{d}^j_{ge}\right\}[\hat{\sigma}_{eg}(t),\hat{\mathcal{O}}(t)]e^{i\omega_{k}(t-t')},
\end{align}   
where $\hat{a}_{k_{F}} = \hat{a}_k(0)e^{-i\omega_k t}$ is the bosonic operator of the freely propagating $k^\text{th}$ field mode. Note that there is a subtlety associated with the substitution of Eq. (\ref{eq6}) into Eq. (\ref{eq5}). Since emitter and field operators commute, the order in which they appear in Eq. (\ref{eq5}) is in principle arbitrary. The emitter part of Eq. (\ref{eq6}), however, does not commute with the emitter operators in Eq. (\ref{eq5}) and the order thus matters when we do the substitution. A more careful treatment shows that the ordering of emitter and field operators in Eq. (\ref{eq5}) is indeed irrelevant provided that a replacement similar to Eq. (\ref{eq6}) is performed for the emitter operators \cite{barnettbook}. In writing Eq. (\ref{eq5}) we have ensured that all expressions are normal ordered such that annihilation operators are to the right and creation operators are to then left. This ensures that these additional terms vanish if the surrounding reservoir is in vacuum. The results derived here are thus only valid in this case and, e.g., not in a thermally excited reservoir where stimulated emission would lead to a modified decay rates. 

Next we use the standard Green's tensor definition in terms of the generalized mode functions $\vec{F}_{k}(\vec{r}_{j})$,
\bea
\label{eq8}
\overleftrightarrow{G}(\vec{r},t,\vec{r'},t') = \sum_{k}\vec{F}_{k}(\vec{r})\vec{F}^\ast_{k}(\vec{r'})e^{-i\omega_{k}(t-t')}
\eea
in Eq. (\ref{eq7}) and use the relation \cite{Novobook06}
\bea
\label{eq9}
\sum_{k}\omega_k \vec{F}_k(\vec{r'})\vec{F}^\ast_k(\vec{r})~e^{-i\omega_k(t-t')} & = & 2\int~d\omega~e^{-i\omega(t-t')} \frac{\omega^2}{\pi c^2}\nonumber\\
&\times&\textbf{Im}\{\overleftrightarrow{G}(\vec{r'},\vec{r},\omega)\},
\eea
where $\overleftrightarrow{G} = \overleftrightarrow{G}(\vec{r'},\vec{r},\omega)$ is the Fourier transform of the Green's tensor corresponding to the emitter-field coupling at some central frequency $\omega$ of the field while $\textbf{Im}\overleftrightarrow{G}(\vec{r'},\vec{r},\omega)$ stands for imaginary part of the Green's tensor. Inserting this in Eq. (\ref{eq7}) we obtain
\begin{widetext}
\bea
\label{eq10}
&&\dot{\hat{\mathcal{O}}} = \frac{i}{\hbar}[\hat{\mathcal{H}}_{c}, \hat{\mathcal{O}}] +\sum_{k}\sum_{j,e,g}\sqrt{\frac{\omega_k}{2\hbar}}\left\{\vec{d}^{j}_{eg}\cdot\vec{F}_{k}(\vec{r}_j)\right\}[\hat{\sigma}_{ge}(t),\hat{\mathcal{O}}(t)]\hat{a}_{k_{F}}+\int~d\omega \left(\frac{\omega^2}{\hbar\pi c^2}\right) \sum_{j,e,g}\sum_{j',e',g'}\int^{t}_{-\infty}dt'\times e^{-i\omega(t-t')} \nonumber\\
&&[\hat{\sigma}_{ge}(t),\hat{\mathcal{O}}(t)]\left\{\vec{d}^{j}_{eg}\cdot\textbf{Im}\overleftrightarrow{G}(\vec{r}_j,\vec{r}_{j'},\omega)\cdot\vec{d}^{j'}_{g'e'}\right\}\hat{\sigma}_{e'g'}(t')-\sum_{k,j,e,g}\sqrt{\frac{\omega_k}{2\hbar}}\left\{\vec{F}^\ast_{k}(\vec{r}_j)\cdot\vec{d}^j_{ge}\right\}\hat{a}^{\dagger}_{k_{F}}[\hat{\sigma}_{eg}(t),\hat{\mathcal{O}}(t)]-\int~d\omega \left(\frac{\omega^2}{\hbar\pi c^2}\right)\nonumber\\
&& \sum_{j,e,g}\sum_{j',e',g'}\int^{t}_{-\infty}dt' e^{i\omega(t-t')}\hat{\sigma}_{g'e}(t')\left\{\vec{d}^{ j'}_{e'g'}\cdot\textbf{Im}\overleftrightarrow{G}(\vec{r}_{j'}, \vec{r}_{j},\omega)\cdot\vec{d}^{j}_{ge}\right\}[\hat{\sigma}_{eg}(t),\hat{\mathcal{O}}(t)],
\eea
\end{widetext}
To solve the time integrals in the above equation we take a closer look at the time-dependent operators and the commutators. The operators are oscillating at the emitter's dipole transition frequencies which is very fast and thus cannot be integrated over time straightaway. Instead we first go to a new frame rotating with the transition frequency to get rid of the fast-oscillating behaviour of the operators and then perform a Markov approximation to get 
\bea
\label{eq11}
\hat{\sigma}_{e'g'}(t') &\approx& \hat{\sigma}_{e'g'}(t)e^{i\omega_{e'g'}(t-t')},
\eea
where $\omega_{e'g'}$ is the frequency of the transition $|e'\rangle \leftrightarrow |g'\rangle$. Note that the Markov approximation accounts for the fact that the timescale over which we consider the emitter dynamics is much larger than any bath correlation times, i.e. we implicitly assume that in the rotating frame the states $|e'\rangle$ and $|g'\rangle$ evolve slowly compared to the bath \cite{Noisebook}. This means that if there is any strong dynamics, e.g., strong dipole-dipole interactions included in $\mathcal{H}_c$ of a strength comparable to the bath correlation time, $|e'\rangle$ and $|g'\rangle$ should be chosen to be the appropriate eigenstates of that interactions. In this situation we can ignore any memory effects of the bath on the emitter dynamics. This mathematically amounts to converting the time dependence $t'$ of the operators to $t$ and replacing the lower bound of the time integral by $-\infty$. Then substituting Eq. (\ref{eq11}) into Eq. (\ref{eq10}) we get
\begin{widetext}
\bea
\label{eq12}
&&\dot{\hat{\mathcal{O}}} = \frac{i}{\hbar}[\hat{\mathcal{H}}_{c}, \hat{\mathcal{O}}] +\sum_{k,j,e,g}\sqrt{\frac{\omega_k}{2\hbar}}\left\{\vec{d}^{j}_{eg}\cdot\vec{F}_{k}(\vec{r}_j)\right\}[\hat{\sigma}_{ge},\hat{\mathcal{O}}]\hat{a}_{k_{F}}+\int d\omega \left(\frac{\omega^2}{\hbar\pi c^2}\right)\sum_{j,e,g}\sum_{j',e',g'}\int^{\infty}_{0}d\tau~e^{-i(\omega-\tilde\omega_{e'g'})\tau}[\hat{\sigma}_{ge},\hat{\mathcal{O}}]\nonumber\\
&&\left\{\vec{d}^{j}_{eg}\cdot\textbf{Im}\overleftrightarrow{G}(\vec{r}_{j},\vec{r}_{j'},\omega)\cdot\vec{d}^{j'}_{g'e'}\right\}\hat{\sigma}_{e'g'}- \sum_{k}\sum_{j,e,g}\sqrt{\frac{\omega_k}{2\hbar}}\left\{\vec{F}^\ast_{k}(\vec{r}_j)\cdot\vec{d}^j_{ge}\right\}\hat{a}^{\dagger}_{k_F}[\hat{\sigma}_{eg},\hat{\mathcal{O}}]-\int d\omega \left(\frac{\omega^2}{\hbar\pi c^2}\right)\sum_{j,e,g}\sum_{j',e',g'}\int^{\infty}_{0}d\tau\nonumber\\
&&~e^{i(\omega-\omega_{e'g'})\tau}\hat{\sigma}_{g'e'}\left\{\vec{d}^{ j'}_{e'g'}\cdot\textbf{Im}\overleftrightarrow{G}(\vec{r}_{j'},\vec{r}_{j},\omega)\cdot\vec{d}^{j}_{ge}\right\}[\hat{\sigma}_{eg},\hat{\mathcal{O}}],
\eea
where $\tau = (t-t')$. 
The time integrals in Eq. (\ref{eq12}) can be expanded into a delta function and a principal value integral to give
\bea
\label{eq13}
&&\dot{\hat{\mathcal{O}}}(t) = \frac{i}{\hbar}[\hat{\mathcal{H}}_{c}, \hat{\mathcal{O}}] +\sum_{k,j,e,g}\sqrt{\frac{\omega_k}{2\hbar}}\left\{\vec{d}^{j}_{eg}\cdot\vec{F}_{k}(\vec{r}_j)\right\}[\hat{\sigma}_{ge},\hat{\mathcal{O}}]\hat{a}_{k_{F}}+\sum_{j,e,g}\sum_{j',e',g'}\int d\omega \left(\frac{\omega^2}{\hbar\pi c^2}\right)\bigg[\pi\delta(\omega-\omega_{e'g'})\nonumber\\
&&- i\textbf{P}\frac{1}{(\omega-\omega_{e'g'}+i\delta\omega)}\bigg][\hat{\sigma}_{ge},\hat{\mathcal{O}}]\bigg\{\vec{d}^{j}_{eg}\cdot\textbf{Im}\overleftrightarrow{G}(\vec{r}_{j},\vec{r}_{j'},\omega)\cdot\vec{d}^{j'}_{g'e'}\bigg\}\hat{\sigma}_{e'g'}-\sum_{k}\sum_{j,e,g}\sqrt{\frac{\omega_k}{2\hbar}}\left\{\vec{F}^\ast_{k}(\vec{r}_j)\cdot\vec{d}^{j}_{ge}\right\}\hat{a}^{\dagger}_{k_F}[\hat{\sigma}_{eg},\hat{\mathcal{O}}]\nonumber\\
&&-\sum_{j,e,g}\sum_{j',e',g'}\int d\omega \left(\frac{\omega^2}{\hbar\pi c^2}\right)\bigg[\pi\delta(\omega-\omega_{e'g'})-i\textbf{P}\frac{1}{(\omega-\omega_{e'g'}+i\delta\omega)}\bigg]\hat{\sigma}_{g'e'}\left\{\vec{d}^{j'}_{e'g'}\cdot\textbf{Im}\overleftrightarrow{G}(\vec{r}_{j'},\vec{r}_{j},\omega)\cdot\vec{d}^{j}_{ge}\right\}[\hat{\sigma}_{eg},\hat{\mathcal{O}}].
\eea
\end{widetext}
On performing the frequency integral and defining
\bea
\label{eq14a}
\Gamma^{jj',ee'}_{g'g} & = &\frac{2(\omega_{e'g'})^{2}}{\hbar c^2}\left\{\vec{d}^{j}_{eg}\cdot\mathbf{Im}\overleftrightarrow{G}(\vec{r}_{j},\vec{r}_{j},\omega_{e'g'})\cdot\vec{d}^{j'}_{g'e'}\right\},\nonumber\\
\\
\label{eq14b}
\Omega^{jj',ee'}_{g'g} & = &\mathbf{P}\int d\omega \left(\frac{\omega^2}{\hbar\pi c^2}\right)\bigg\{\frac{\vec{d}^{j}_{eg}\cdot\mathbf{Im}\overleftrightarrow{G}\cdot\vec{d}^{j'}_{g'e'}}{(\omega-\omega_{eg}+i\delta\omega)}\bigg\},
\eea
we get the Heisenberg equation of motion for any arbitrary operator $\mathcal{O}$ acting on the emitters, 
\bea
\label{eq15}
\dot{\hat{\mathcal{O}}} & = &\frac{i}{\hbar}[\hat{\mathcal{H}}_{c}, \hat{\mathcal{O}}] +\sum_{k}\sum_{j,e,g}\sqrt{\frac{\omega_k}{2\hbar}}\bigg(\left\{\vec{d}^{j}_{eg}\cdot\vec{F}_{k}(\vec{r}_j)\right\}\nonumber\\
&\times&[\hat{\sigma}_{ge},\hat{\mathcal{O}}]\hat{a}_{k_{F}}-\left\{\vec{F}^\ast_{k}(\vec{r}_j)\cdot\vec{d}^{j}_{ge}\right\}\hat{a}^{\dagger}_{k_{F}}[\hat{\sigma}_{eg},\hat{\mathcal{O}}]\bigg)\nonumber\\
&+&\sum_{j,j'}\sum_{e,g}\sum_{e',g'}[\hat{\sigma}_{ge},\hat{\mathcal{O}}]\left(\frac{1}{2}\Gamma^{jj',ee'}_{g'g}-i\Omega^{jj',ee'}_{g'g}\right)\hat{\sigma}_{e'g'}\nonumber\\
&-&\sum_{j,j'}\sum_{e,g}\sum_{e',g'}\hat{\sigma}_{g'e'}\left(\frac{1}{2}\Gamma^{j'j,e'e}_{gg'}+i\Omega^{j'j,e'e}_{gg'}\right)[\hat{\sigma}_{eg},\hat{\mathcal{O}}].\nonumber\\
\eea

Note that in Eq. (\ref{eq14a}) we have derived the general expression for the decay rate $\Gamma$ due to field mediated interferences from emitters located at positions $\vec{r}_{j}$ and $\vec{r}_{j'}$. For $j' = j$ the diagonal terms $\Gamma^{j,e}_{g}$ involves dipole moments of the same transition and thus corresponds to the total spontaneous decay rate of a single emitter. The off-diagonal terms signify induced dynamics and describes the collective decay of the emitters coupled to the same vacuum reservoir. Furthermore, $\Omega^{jj', ee'}_{g'g}$ in Eq. (\ref{eq14b}) contributes to the field-mediated (dipolar) shift of the energy levels of the multi-level emitters. For $j = j'$ this is similar to the Lamb shift of a single-emitter energy level induced by the vacuum. As noted before, this expression is derived within the rotating-wave approximation and thus does not correctly describe the dipolar interaction between emitters separated by less than a wavelength. Had we not performed the rotating-wave approximation, the expression for $\Omega$ would have an additional term with the denominator of the Green's function being $(\omega+\omega_{eg}+i\delta\omega)$. For emitters separated by more than a wavelength the interaction is dominated by the terms fulfilling energy conservation (or resonant scattering) which then leads to the same $\Omega$ that we derived in Eq. (\ref{eq14b}). For applications with nearby emitters our expression in Eq. (\ref{eq14b}) should be replaced by the correct expression and the formalism below can still be applied. It is worth noting that the above derived equation of motion for the emitters serves as a foundation to many studies in stimulated Raman scattering as well as in coupling between atomic spin excitation and collective emission of light \cite{Ray81,Aker08, Ped09, Por08, Maz07, Mar09, Anders10}. 

\section{Solving for the coherence}
In this section using Eq. (\ref{eq15}) we find solutions to the operator equations corresponding to the coherences of the emitter system \cite{Scullyb}. From Eq. (\ref{eq15}) we find the equation for the coherence operator 
\bea
\label{eq16}
\dot{\hat{\sigma}}_{eg} & = &\frac{i}{\hbar}[\hat{\mathcal{H}}_{c}, \hat{\sigma}_{eg}] +\sum_{k}\sum_{j}\sum_{e',g'}\sqrt{\frac{\omega_k}{2\hbar}}\nonumber\\
&\times&\left\{\vec{d}^{j}_{e'g'}\cdot\vec{F}_{k}(\vec{r}_j)\right\}[\hat{\sigma}_{g'e'},\hat{\sigma}_{eg}]\hat{a}_{k_{F}}+\sum_{jj'}\sum_{e'g'}\nonumber\\
&&\sum_{e''g''}[\hat{\sigma}_{g'e'},\hat{\sigma}_{eg}]\left(\frac{1}{2}\Gamma^{jj'e'e''}_{g''g'}-i\Omega^{jj'e'e''}_{g''g'}\right)\hat{\sigma}_{e''g''}.\nonumber\\
\eea
A closer look at the commutator of $\hat{\sigma}_{eg}$ and the bare Hamiltonian $\mathcal{H}_{c}$ suggests that on performing the commutation we get back a similar coherence operator since $\mathcal{H}_{c}$ does not couple $M_{e}$ and $M_{g}$. Similarly, 
on expanding the term $[\hat{\sigma}_{g'e'},\hat{\sigma}_{eg}]\hat{\sigma}_{e''g''}$ we find that the last term in the above equation becomes
\bea
\label{eq17}
-\sum_{e'g'}\left(\frac{1}{2}\Gamma^{jj'ee'}_{g' g'}-i\Omega^{jj'e e'}_{g' g'}\right)\hat{\sigma}_{e'g}, 
\eea 
where we have renamed the index $e''$ as $e'$. Note that $\sigma_ {eg}$ is not only coupled to itself $e'=e$ but also to the coherences containing all other excited state $e' \neq e$.

We can reduce the operator equation in Eq. (\ref{eq16}) to the form
\bea
\label{eq18}
\dot{\hat{\sigma}}_{eg} &=& -\frac{i}{\hbar}\sum_{e'}(\mathcal{H}_{nh})_{ee'}\hat{\sigma}_{e'g}+\sum_{k}\sqrt{\frac{\omega_k}{2\hbar}}\bigg(\sum_{j,e'}\bigg\{\vec{d}^{j}_{e'g}\nonumber\\
&&\cdot\vec{F}_{k}(\vec{r}_{j})\bigg\}\hat{\sigma}_{ee'}-\sum_{j,g'}\hat{\sigma}_{g'g}\left\{\vec{d}^{j}_{e g'}\cdot\vec{F}_{k}(\vec{r}_{j})\right\}\bigg)\hat{a}_{k_{F}},\nonumber\\
\eea
where we have introduced the ground state operator $\hat{\sigma}_{g'g} = |g\rangle\langle g'|$ defined in the ground state subspace $M_{g}$. Again note the order of the indices: similar to the definition of $\sigma_{ge}$ this order ensures that we can find the elements of the density matrix through as $\rho_{ij} = \langle \sigma_{ij}\rangle$. Furthermore, $(\mathcal{H}_{nh})_{ee'} = \langle e|\hat{\mathcal{H}}_{nh}|e'\rangle$ is the matrix element of the non-Hermitian Hamiltonian $\mathcal{H}_{nh}$ in the basis of the excited states that span the excited state subspace $M_e$. The non-Hermitian Hamiltonian $\hat{\mathcal{H}}_{nh}$ is defined as
\bea
\label{eq19}
\hat{\mathcal{H}}_{nh}= \hat{\mathcal{H}}_{c_{e}}-i\sum_{jj'}\sum_{g'}\left(\frac{1}{2}\Gamma^{jj' ee'}_{g' g'}-i\Omega^{jj'e e'}_{g' g'}\right)\sigma_{e'e}.\nonumber\\
\eea
This non-Hermitian Hamiltonian is well known in the theory of Monte-Carlo wavefunctions where it describes the evolution of the system in the absence of decay \cite{Dali92}. We will see later that this non-Hermitian Hamiltonian is central to the dynamics of the multi-emitter systems. It is hence important to understand the individual contributions in Eq. (\ref{eq19}). The Hamiltonian $\hat{\mathcal{H}}_{c_{e}}$ consists of terms corresponding to the excited state energies along with any intra- and inter-emitter interactions that are not mediated by the field in the dielectric medium. Importantly for the formalism to be applicable, this term should only act within the excited subspace and should not contain any coupling between the ground and excited states. Furthermore, it is worth emphasizing that, $\hat{\mathcal{H}}_{c_{e}}$ also allows the formalism to deal with complex intra- and inter-emitter dynamics. The term in the bracket in Eq. (\ref{eq19}) for $j \neq j'$ corresponds to the collective decay and energy shift of the emitters as explained before. For $j = j'$ the first terms in the round bracket corresponds to $\Gamma_{e}$, the total natural linewidth of an excited state $|e\rangle$ for the case of a simple two-level emitter, while for multi-level emitters this may contain decay induced intra-emitter coupling between the excited states. The second term inside the round bracket in Eq. (\ref{eq19}) as discussed before is analogous to a single-emitter Lamb shift $\sim\sum_{g}\Omega^{e}_{g}$ for $j = j'$, $e = e'$ while for $e \neq e'$ it represents excited state couplings due to intra-emitter interference among different dipole transitions pathways. Here we have considered $|e\rangle$ to be a single emitter excited-state. The above discussion will be slightly different if $|e\rangle$ is a collective state. In writing $\hat{\mathcal{H}}_{c_{e}}$ we have neglected the Hamiltonian acting on the ground states. This is justified if it is a perturbation. However if the couplings among the ground states are non-perturbative, one needs to include them in $\hat{\mathcal{H}}_{nh}$. A prescription for doing this was already laid out in Ref. \cite{Reiter12}. In the simple case where the ground state Hamiltonian can be assumed to be diagonal, the non-Hermitian Hamiltonian of the system $\hat{\mathcal{H}}_{nh}$ needs to be substituted by an initial state dependent one given by $\hat{\mathcal{H}}_{nh_{g}} = (\hat{\mathcal{H}}_{nh} - \hat{I}E_{g})$, where $E_{g}$ is the energy of the ground state $|g\rangle$ and $\hat{I}$ is  the identity operator. The Hamiltonain $\hat{\mathcal{H}}_{nh_{g}}$, then gives a suitable description for processes originating from the state $|g\rangle$.

Next we assume that the emitters are initially in the ground state and that the time scale of $\hat{H}_{nh}$ is much faster than the coupling between the ground and excited states. Since $\hat{H}_{nh}$ includes both decay and detuning of the excited-state, this amounts to the standard approximation of adiabatic elimination. Then, using the definition of electric field from Eq. (\ref{eq2}) we can rewrite Eq. (\ref{eq18}) as 
\bea
\label{eq20}
\dot{\hat{\sigma}}_{eg} &=& -\frac{i}{\hbar}\sum_{e'}(\mathcal{H}_{nh})_{ee'}\hat{\sigma}_{e'g}+i\sum_{jg'}\left\{\vec{d}^{j}_{eg'}\cdot\frac{\mathcal{E}^{+}(\vec{r}_{j},t)}{\hbar}\right\}\hat{\sigma}_{g'g},\nonumber\\
\eea
the solution of which is given by
\bea
\label{eq21}
\hat{\sigma}_{eg} &=&i\sum_{jg'}\int^{t}_{-\infty}dt' \langle e|\exp[-i\sum_{e'}\tilde{\mathcal{H}}_{nh}(t-t')]|e'\rangle\nonumber\\
&\times&\left\{\vec{d}^{j}_{e'g'}\cdot\frac{\mathcal{E}^{+}(\vec{r}_{j})}{\hbar}\right\}\hat{\sigma}_{g'g}.\nonumber\\
\eea 
Here $\tilde{\mathcal{H}}_{nh} = (\mathcal{H}_{nh}-\hbar\omega)/\hbar$ with $\omega = \omega_{k}(k_{0})$, $k_0$ being the central wavenumber corresponding to the incoming photon. In evaluating the coherence $\hat{\sigma}_{eg}$ above we have invoked our primary assumption of the incident light field being weak. We can then neglect any two-photon scattering processes from the system of emitters and all higher order excitation terms involving solely the excited states like $\hat{\sigma}_{ee}$. Thus the equation for coherence in (\ref{eq18}) can be approximated to contain terms involving only the ground states and single excitation of the emitters as given explicitly by Eq. (\ref{eq21}). Furthermore, we have assumed that the ground state operator $\hat{\sigma}_{g' g}$ is slowly varying so that it is effectively a constant over the period of integration. 

On solving the integral in Eq. (\ref{eq21}) we then get 
\bea
\label{eq22}
\hat{\sigma}_{eg} &=&\sum_{jg'}\sum_{e'}[\tilde{\mathcal{H}}_{nh}]^{-1}_{ee'}\left\{\vec{d}^{j}_{e'g'}\cdot\frac{\mathcal{E}^{+}(\vec{r}_{j})}{\hbar}\right\}\hat{\sigma}_{g'g}.
\eea 
From Eq. (\ref{eq21}) we see that the solution of the coherence for the emitters can be found in terms of the inverse of the non-Hermitian Hamiltonian $\tilde{\mathcal{H}}_{nh}$, and the dynamics of the ground states. Thus, our formalism can deal with any possible complex dynamics, as long as the non-Hermitian Hamiltonian can be inverted. Note that in the simplest case of a two level system with a single optical transition, $\tilde{\mathcal{H}}^{-1}_{nh}$ is just $1/(\Delta-i\Gamma/2)$, where $\Delta$ is the detuning of the incoming light from the transition frequency and $\Gamma$ is the total decay rate corresponding to the transition. In a later section we will develop a master equation to find the solution of the ground state dynamics. In the following section, however, we focus on investigating the photon scattering dynamics from the system of emitters. 
\section{Scattering dynamics of the incident photons}
Now that we have the solution for the coherences we use it to develop the key result of this work, a scattering relation between the input and scattered field for a generic multi-emitter system in any dielectric medium. For this purpose we substitute Eq. (\ref{eq22}) into Eq. (\ref{eq6}) to obtain
\bea
\label{eq23}
&&\hat{a}_{k}(t)=\hat{a}_{k_{F}}+\sum_{jj'}\sum_{gg'}\sqrt{\frac{\hbar\omega_{k}}{2\epsilon}}\int^{t}_{-\infty} dt e^{i(\omega_{gg'}-\omega_{k})(t-t')}\nonumber\\
&&\times\sum_{ee'}\left\{\frac{\vec{F}^\ast_{k}(\vec{r}_{j})\cdot\vec{d}^{j}_{ge}}{\hbar}\right\}\left[\tilde{\mathcal{H}}_{nh}\right]^{-1}_{ee'}\left\{\frac{\vec{d}^{j'}_{e'g'}\cdot\vec{\mathcal{E}}^{+}(\vec{r}_{j'})}{\hbar}\right\}\hat{\sigma}_{g'g},\nonumber\\
\eea
where we have used the slowly varying nature of the ground-state operators to write $\hat{\sigma}_{g'g}(t') = \hat{\sigma}_{g'g}(t)e^{-i\omega_{gg'}(t-t')}$ with $\omega_{gg'} = (\omega_{g}-\omega_{g'})$ being the frequency difference between the ground states. 

To convert Eq. (\ref{eq23}) into an equation for the quantized electric field at a certain position in space and time, we sum over the field-mode operators in the form  i$\sum_{k}\sqrt{\frac{\hbar\omega_{k}}{2}}\vec{F}_{k}(\vec{r})\hat{a}_{k}$ and use the relationship $i\sqrt{\frac{\hbar\omega_{k'}}{2}}\hat{a}_{k'}(0) = \int d\vec{r^\prime} \vec{F}^{\dagger}_{k'}(\vec{r^\prime})\epsilon(\vec{r^\prime})\hat{\vec{\mathcal{E}}}^{+}(\vec{r^\prime},0)$ along with the Green's function expansion of the field to get 
\bea
\label{eq24}
\hat{\vec{\mathcal{E}}}^{+}(\vec{r},t) & = &\int d\vec{r^\prime}~\epsilon(\vec{r^\prime})\mathbf{G}(\vec{r},t,\vec{r^\prime},0)\hat{\vec{\mathcal{E}}}^{+}(\vec{r^\prime},0)+\left(\frac{i\omega}{2\hbar}\right)\nonumber\\
&\times&\sum_{jj'}\sum_{gg'}\int^{t}_{-\infty} dt' ~e^{i\omega_{gg'}(t-t')}\hat{\sigma}_{g'g} (t)\mathbf{G}(\vec{r},t,\vec{r}_{j},t')\nonumber\\
&\times&\sum_{ee}\bigg(\vec{d}^{j}_{ge}[\tilde{\mathcal{H}}_{nh}]^{-1}_{ee'}\vec{d}^{j'}_{e'g'}\bigg)\int d\vec{r^\prime}~\epsilon(\vec{r^\prime})\nonumber\\
&\times&\mathbf{G}(\vec{r}_{j'},t',\vec{r^{\prime}},0)\hat{\vec{\mathcal{E}}}^{+}(\vec{r^\prime},0).
\eea
Here, $\vec{r}$ is the point of observation, $\vec{r^\prime}$ is some initial spatial position of the incident field while $\vec{r}_{j}$ and $\vec{r}_{j'}$ corresponds to the spatial position of emitters $j$ and $j'$ respectively. The first term on the right hand side of Eq. (\ref{eq23}) represents the freely propagating field with the Green's function being simply a propagator. The second term represents the scattering event and gives the scattered field including the dynamical response of the emitters. 

We next expand the Green's function in Eq. (\ref{eq24}) in terms of the mode functions $\vec{F}_{k}$, and evaluate the space integral using the orthogonality condition in Eq. (\ref{eq3}). Furthermore, on doing the Fourier transform from the time to frequency domain and defining the input field as $\hat{\vec{\mathcal{E}}}_{in}(\vec{r},t) = \sum_{k}\sqrt{\frac{\hbar\omega_{k}}{2}}\vec{F}_{k}(\vec{r})\hat{a}_{k}(0)e^{-i\omega_{k}t}$, Eq. (\ref{eq24}) can be transformed into 
\bea
\label{eq25}
\hat{\vec{\mathcal{E}}}^{+}(\vec{r},t) & = &\hat{\vec{\mathcal{E}}}_{in}(\vec{r},t)+\left(\frac{i\omega}{2\hbar}\right)\sum_{jj'}\sum_{gg'}\mathbf{G}(\vec{r},\vec{r}_{j},\omega-\omega_{gg'})\nonumber\\
&\times&\hat{\sigma}_{g'g}\sum_{e}\bigg(\vec{d}^{j}_{ge}[\tilde{\mathcal{H}}_{nh}]^{-1}_{ee'}\vec{d}^{j'}_{e'g'}\bigg)\hat{\vec{\mathcal{E}}}_{in}(\vec{r}_{j'},t).
\eea
Here $\mathbf{G}(\vec{r},\vec{r}_{j},\omega-\omega_{gg'})$, the Fourier transform of the Green's tensor, gives the response of the field corresponding to the characteristic frequency $(\omega-\omega_{gg'})$ of the dielectric medium containing the emitters. 
 
Eq. (\ref{eq25}) formulates a scattering relation between a weak input field and the output scattered field from a system of emitters in some dielectric medium and is the key result of this paper. Note that Eq. (\ref{eq25}) has the following salient features: (a) it gives complete solution of the scattering problem considering Markovian dynamics, (b) it includes emitters that are fully generic and (c) it deals with the completely general case of several multi-level emitters coupled to the field in some dielectric medium. 

It important to note that in the derived photon scattering relation all the system properties are included through the non-Hermitian Hamiltonian $\tilde{\mathcal{H}}_{nh}$ while the dynamical evolution of the emitters is through the evolution of the ground states. Thus, to get the complete photon scattering dynamics we need to find the solution of the ground state dynamics. This can be quite challenging depending on the complexity of the system. In the next section however, by exploiting the formulation of EOs \cite{Reiter12}, which again involves the inverse of the non-Hermitian Hamiltonian $[\tilde{\mathcal{H}}_{nh}]^{-1}$, we find a master equation for such ground-state evolution. It is then a simple algebraic/numerical exercise to solve the master equation depending on the size of the Hilbert space of the emitters.  

Given the generic nature of Eq. (\ref{eq25}), in principle it can be applicable to any dielectric medium for which one can calculated the Green's function. For example, in case of the $1$D waveguide, the Green's function in Eq. (\ref{eq25}) is quite straightforward, and one can get a simple photon-scattering relation between the input and output mode. In the following paper \cite{Dasp2}, we consider these cases and show explicitly how Eq. (\ref{eq25}) can be utilized to directly achieve the photon reflected and transmitted amplitude in numerous problems, which are otherwise non-trivial to achieve via other methods.
\section{Effective operator master equation for the ground state dynamics}
In this section we develop a master equation to solve the ground state evolution for the emitters in the Heisenberg picture. The master equation will be derived in the effective operator formalism in a spirit similar to that of Ref.  \cite{Reiter12}. For this purpose we first write down the equation of motion for the ground state operator $\hat{\sigma}_{g'g}$ using Eq. (\ref{eq15}) as
\bea
\label{eq26}
\dot{\hat{\sigma}}_{g'''g''} & = &\frac{i}{\hbar}[\hat{\mathcal{H}}_{c}, \hat{\sigma}_{g'''g''}] +\sum_{k}\sum_{j}\sum_{eg}\sqrt{\frac{\omega_k}{2\hbar}}\nonumber\\
&&\bigg(\left\{\vec{d}^{j}_{eg}\cdot\vec{F}_{k}(\vec{r}_j)\right\}[\hat{\sigma}_{ge},\hat{\sigma}_{g'''g''} ]\hat{a}_{k_{F}}-\left\{\vec{F}^\ast_{k}(\vec{r}_j)\cdot\vec{d}^{j}_{ge}\right\}\nonumber\\
&&\hat{a}^{\dagger}_{k_{F}}[\hat{\sigma}_{eg},\hat{\rho}_{g'''g''} ]\bigg)+\sum_{j,j'}\sum_{e,g}\sum_{e',g'}[\hat{\sigma}_{ge},\hat{\sigma}_{g'''g''}]\nonumber\\
&&\bigg(\frac{1}{2}\Gamma^{jj',ee'}_{g'g}-i\Omega^{jj',ee'}_{g'g}\bigg)\hat{\sigma}_{e'g'}-\sum_{j,j'}\sum_{e,g}\sum_{e',g'}\nonumber\\
&&\hat{\sigma}_{g'e'}\bigg(\frac{1}{2}\Gamma^{j'j,e'e}_{gg'}+i\Omega^{j'j,e'e}_{gg}\bigg)[\hat{\sigma}_{eg},\hat{\sigma}_{g'''g''}].
\eea
The commutator in the $4^\text{th}$ and $5^\text{th}$ term of the above equation on evaluation gives us, respectively, 
\bea
\label{eq27}
&&\sum_{jj'}\sum_{ee'}\sum_{g'}\hat{\sigma}_{g'''e}\hat{\sigma}_{e'g'}\left(\frac{1}{2}\Gamma^{jj'ee'}_{g'g''}-i\Omega^{jj'ee'}_{g'g''}\right),\\
\label{eq28}
&-&\sum_{jj'}\sum_{ee'}\sum_{g'}\left(\frac{1}{2}\Gamma^{j'je'e}_{g''' g'}+i\Omega^{j'je'e}_{g''' g'}\right)\hat{\sigma}_{g'e'}\hat{\sigma}_{eg''}. 
\eea
Using  Eq. (\ref{eq22}) we substitute for $\hat{\sigma}_{g'e'}$ and $\hat{\sigma}_{g''e}$ and their Hermitian conjugates in Eqs. (\ref{eq27}) and (\ref{eq28}). After some tedious algebra we get the equation of motion of the ground state operator as
\begin{widetext}
\bea
\label{eq29}
&&\dot{\hat{\sigma}}_{g'''g''} =  \frac{i}{\hbar}[\mathcal{H}_{c},\hat{\sigma}_{g'''g''}]-i\sum_{jj'}\sum_{ee'}\sum_{g'}\hat{\sigma}_{g'''g'}\bigg(\vec{d}^{j'}_{g'e'}\cdot\frac{\mathcal{\vec{E}}^{-}(\vec{r}_{j'})}{\hbar}\bigg)[\tilde{\mathcal{H}}^{\dagger}_{nh}]^{-1}_{e'e}\bigg(\vec{d}^{j}_{eg''}\cdot\frac{\vec{\mathcal{E}}^{+}(\vec{r}_{j})}{\hbar}\bigg)+i\sum_{jj'}\sum_{ee'}\sum_{g'}\nonumber\\
&&\bigg(\frac{\mathcal{\vec{E}}^{-}(\vec{r}_{j})}{\hbar}\cdot\vec{d}^{j}_{g'''e}\bigg)[\tilde{\mathcal{H}}_{nh}]^{-1}_{ee'}\bigg(\vec{d}^{j'}_{e'g'}\cdot\frac{\mathcal{\vec{E}}^{+}(\vec{r}_{j'})}{\hbar}\bigg)\hat{\sigma}_{g'g''}+\sum_{jj'}\sum_{j''j'''}\sum_{ee'}\sum_{e'' e'''}\sum_{g''''}\sum_{g'}\hat{\sigma}_{g'''g''''}\bigg\{\vec{d}^{j''}_{g''''e''}\cdot\frac{\vec{\mathcal{E}}^{-}(\vec{r}_{j''})}{\hbar}\bigg\}[\tilde{\mathcal{H}}^\dagger_{nh}]^{-1}_{e''e}\nonumber\\
&&\Gamma^{j'j e e'}_{g'g''}[\tilde{\mathcal{H}}_{nh}]^{-1}_{e'e'''}\sum_{g'''''}\bigg\{\vec{d}^{j'''}_{e'''g'''''}\cdot\frac{\vec{\mathcal{E}}^{+}(\vec{r}_{j'''})}{\hbar}\bigg\}\hat{\sigma}_{g'''''g'}.
\eea 
where we have used the notation $\mathcal{\vec{E}}^{\pm}(\vec{r}_{l}) = \mathcal{\vec{E}}^{\pm}_{l}$. 
On using $\hat{\sigma}_{g'''g''''}\hat{\sigma}_{g'''''g'} = \hat{\sigma}_{g'''''g''''}\delta_{g'g'''}$ in Eq. (\ref{eq29}) and rearranging we can write Eq. (\ref{eq29}) as
\bea
\label{eq31}
&&\dot{\hat{\sigma}}_{g'g} =  \frac{i}{\hbar}[\mathcal{H}_{c},\hat{\sigma}_{g'g}]-i\sum_{jj'}\sum_{ee'}\sum_{g''}\hat{\sigma}_{g'g''}\left(\vec{d}^{j'}_{g''e'}\cdot\frac{\mathcal{\vec{E}}^{-}(\vec{r}_{j'})}{\hbar}\right)[\tilde{\mathcal{H}}^{\dagger}_{nh}]^{-1}_{e'e}\left(\frac{\vec{d}^{j}_{eg}\cdot\vec{\mathcal{E}}^{+}(\vec{r}_{j})}{\hbar}\right)+i\sum_{jj'}\sum_{ee'}\sum_{g''}\bigg(\frac{\mathcal{\vec{E}}^{-}(\vec{r}_{j})}{\hbar}\cdot\vec{d}^{j}_{g'e}\bigg)\nonumber\\
&&[\tilde{\mathcal{H}}_{nh}]^{-1}_{ee'}\left(\vec{d}^{j'}_{e'g''}\cdot\frac{\mathcal{\vec{E}}^{+}(\vec{r}_{j'})}{\hbar}\right)\hat{\sigma}_{g''g}+\sum_{jj'}\sum_{j''j'''}\sum_{ee'}\sum_{e'' e'''}\sum_{g''g'''}\hat{\sigma}_{g'''g''}\bigg\{\vec{d}^{j''}_{g''e''}\cdot\frac{\vec{\mathcal{E}}^{-}(\vec{r}_{j''})}{\hbar}\bigg\}[\tilde{\mathcal{H}}^\dagger_{nh}]^{-1}_{e''e}\Gamma^{j'j e e'}_{g'g}[\tilde{\mathcal{H}}_{nh}]^{-1}_{e'e'''}.\nonumber\\
&&\bigg\{\vec{d}^{j'''}_{e'''g'''}\cdot\frac{\vec{\mathcal{E}}^{+}(\vec{r}_{j'''})}{\hbar}\bigg\}
\eea 
\end{widetext}
Note that, in writing Eq. (\ref{eq31}), for notational convenience we have done the following renaming $(g'''= g', g''= g, g'= g'', g''''= g'', g'''''= g''')$, as the primes over $g$ are simple floating indices.

Following Ref. \cite{Reiter12}, we next define a perturbative excitation (de-excitation) operator
\bea
\label{eq33}
\hat{\mathcal{A}}^{j+}_{eg} = \left\{\vec{d}^{j}_{eg}\cdot\frac{\mathcal{\vec{E}}^{+}(\vec{r}_{j})}{\hbar}\right\}, 
\eea
with the Hermitian conjugate $\hat{\mathcal{A}}^{j-}_{eg}= [\hat{\mathcal{A}}^{j+}_{ge}]^{\dagger}$. Using Eq. (\ref{eq33}) and its Hermitian conjugate in Eq. (\ref{eq31}) we get,
\bea
\label{eq34}
\dot{\hat{\sigma}}_{g'g} & = & \frac{i}{\hbar}[\mathcal{H}_{c},\hat{\sigma}_{g'g}]-i\sum_{jj'}\sum_{ee'}\sum_{g''}\hat{\sigma}_{g'g''}\hat{\mathcal{A}}^{j'-}_{g''e'}[\tilde{\mathcal{H}}^{\dagger}_{nh}]^{-1}_{e'e}\nonumber\\
&&\hat{\mathcal{A}}^{j+}_{eg}+i\sum_{jj'}\sum_{ee'}\sum_{g''}\hat{\mathcal{A}}^{j-}_{g'e}[\tilde{\mathcal{H}}_{nh}]^{-1}_{ee'}\hat{\mathcal{A}}^{j'+}_{e'g''}\hat{\sigma}_{g''g'}\nonumber\\
&&+\sum_{jj'}\sum_{j''j'''}\sum_{ee'}\sum_{e'' e'''}\sum_{g''g'''}\hat{\sigma}_{g'''g''}\hat{\mathcal{A}}^{j''-}_{g''e''}[\tilde{\mathcal{H}}^\dagger_{nh}]^{-1}_{e''e}\nonumber\\
&&\Gamma^{j'j e e'}_{g'g}[\tilde{\mathcal{H}}_{nh}]^{-1}_{e'e'''}\hat{\mathcal{A}}^{j'''+}_{e'''g'''}.
\eea

From Eq. (\ref{eq14a}) we see that the decay term $\Gamma^{j'jee'}_{gg'}$, is proportional to the imaginary part of the Fourier transform of the Green's tensor $\overleftrightarrow{G}$. The Green's tensor in turn can be expanded in the basis of the orthogonal eigenmode functions $\vec{F}_{k}$ for the electromagnetic field. On noting that $\overleftrightarrow{G}(\vec{r}_{j},\vec{r}_{j'},\omega) = \sum_{k} c^{2}\vec{F}_{k}(\vec{r}_{j},\omega_{k})\vec{F}^\ast_{k}(\vec{r}_{j'},\omega_{k})/(\omega^{2}_{k}-\omega^{2})$ and following the discussion in Ref. \cite{Novobook06} we find that $\mathbf{Im}\overleftrightarrow{G}(\vec{r}_{j},\vec{r}_{j}, \omega) = \sum_{k}(\pi c^{2}/\omega)\vec{F}_{k}(\vec{r}_{j},\omega_{k})\vec{F}^\ast_{k}(\vec{r}_{j'},\omega_{k})\delta(\omega-\omega_{k})$. On substituting this into Eq. (\ref{eq14a}) we find 
\bea
\label{eq34a}
\Gamma^{j'j ee'}_{g'g} = \sum_{k|\omega_{k} = \omega}\frac{2\pi\omega_{k}}{\hbar c}\bigg[\vec{d}^{j}_{eg}\cdot\vec{F}_{k}(\vec{r}_{j})\vec{F}^\ast_{k}(\vec{r}_{j'})\cdot\vec{d}^{j'}_{g'e}\bigg]~
\eea
Here we have assumed that the mode functions $F_{k}$ are slowly varying functions of the frequency $\omega_{k}$. The sum over mode functions then reduces to a sum over the total energy of the modes and a sum over the remaining transverse and polarization degrees of freedom. Note that in deriving Eq. (\ref{eq34a}), the sum over the total energy of the modes has been performed and thus, the summation over $k$ only runs over the transverse modes and polarization degrees of freedom.  Now we introduce a new operator $\hat{c}^{k}_{ij}$ such that $\Gamma^{jj 'ee'}_{gg'}$ is diagonal in the eigenbasis of this operator and can be represented as $\Gamma^{j'j ee'}_{g'g}=\sum_{k} \hat{c}^{k\dagger}_{eg}\hat{c}^{k}_{g'e'}$, where
\bea
\label{eq35}
\hat{c}^{k\dagger}_{eg} & = & \sqrt{\frac{2\pi\omega_{k}}{\hbar}}\sum_{j}\left[\vec{d}^{j}_{eg}\cdot\vec{F}_{k}(\vec{r}_{j})\right],\nonumber\\
\hat{c}^{k}_{g'e'} & = &  \sqrt{\frac{2\pi\omega_{k}}{\hbar}}\sum_{j'}\left[\vec{F}^\ast_{k}(\vec{r}_{j'})\cdot\vec{d}^{j'}_{g'e'}\right].
\eea
These operators are equivalent to the standard jump operators appearing in a master equation of Lindblad form \cite{Scullyb}. 

Substituting Eq. (\ref{eq35}) into Eq. (\ref{eq34}) we get 
\bea
\label{eq36}
\dot{\hat{\sigma}}_{g'g} & = &\frac{i}{\hbar}[\mathcal{H}_{c},\hat{\sigma}_{g'g}]-i\sum_{jj'}\sum_{ee'}\sum_{g''}\hat{\sigma}_{g'g''}\hat{\mathcal{A}}^{j'-}_{g''e'}[\tilde{\mathcal{H}}^{\dagger}_{nh}]^{-1}_{e'e}\nonumber\\
&&\hat{\mathcal{A}}^{j+}_{eg}+i\sum_{jj'}\sum_{ee'}\sum_{g''}\hat{\mathcal{A}}^{j-}_{g'e}[\tilde{\mathcal{H}}_{nh}]^{-1}_{ee'}\hat{\mathcal{A}}^{j'+}_{e'g''}\hat{\sigma}_{g''g'}\nonumber\\
&&+\sum_{jj'}\sum_{ee'}\sum_{e'' e'''}\sum_{g''g'''}\hat{\sigma}_{g'''g''}\hat{\mathcal{A}}^{j-}_{g''e''}[\tilde{\mathcal{H}}^\dagger_{nh}]^{-1}_{e''e}\nonumber\\
&&\sum_{k} \hat{c}^{k\dagger}_{eg}\hat{c}^{k}_{g'e'}[\tilde{\mathcal{H}}_{nh}]^{-1}_{e'e'''}\hat{\mathcal{A}}^{j'+}_{e'''g'''}.
\eea
Re-arranging the terms in Eq. (\ref{eq36}) we get 
\begin{widetext}
\bea
\label{eq37}
\dot{\hat{\sigma}}_{g'g} & = &\frac{i}{\hbar}[\mathcal{H}_{c},\hat{\sigma}_{g'g}]-i\sum_{jj'}\sum_{ee'}\sum_{g''}\hat{\sigma}_{g'g''}\hat{\mathcal{A}}^{j'-}_{g''e'}[\tilde{\mathcal{H}}^{\dagger}_{nh}]^{-1}_{e'e}\hat{\mathcal{A}}^{j+}_{eg}+i\sum_{jj'}\sum_{ee'}\sum_{g''}\hat{\mathcal{A}}^{j-}_{g'e}[\tilde{\mathcal{H}}_{nh}]^{-1}_{ee'}\hat{\mathcal{A}}^{j'+}_{e'g''}\hat{\sigma}_{g''g'}\nonumber\\
&&+\sum_{k} \sum_{jj'}\sum_{ee'}\sum_{e'' e'''}\sum_{g''g'''}\hat{\mathcal{A}}^{j-}_{g''e''}[\tilde{\mathcal{H}}^\dagger_{nh}]^{-1}_{e''e}\hat{c}^{k\dagger}_{eg}\hat{\sigma}_{g'''g''}~\hat{c}^{k}_{g'e'}[\tilde{\mathcal{H}}_{nh}]^{-1}_{e'e'''}\hat{\mathcal{A}}^{j'+}_{e'''g'''}.
\eea
\end{widetext}
Furthermore, we introduced a new operator defined as $\hat{\mathcal{M}}_{gg'} = \sum_{jj'}\sum_{ee'}\hat{\mathcal{A}}^{j'-}_{ge}[\tilde{\mathcal{H}}_{nh}]^{-1}_{ee'}\hat{\mathcal{A}}^{j+}_{e'g'}$ in Eq. (\ref{eq37}) to get,
\bea
\label{eq38}
\dot{\hat{\sigma}}_{g'g} & = &\frac{i}{\hbar}[\mathcal{H}_{c},\hat{\sigma}_{g'g}]-i\sum_{g''}\hat{\sigma}_{g'g''}\hat{\mathcal{M}}^\dagger_{g''g}+i\sum_{g''}\hat{\mathcal{M}}_{g'g''}\hat{\sigma}_{g''g}\nonumber\\
&&+\sum_{k} \sum_{jj'}\sum_{ee'}\sum_{e'' e'''}\sum_{g''g'''}\hat{\mathcal{A}}^{j-}_{g''e''}[\tilde{\mathcal{H}}^\dagger_{nh}]^{-1}_{e''e}\hat{c}^{k\dagger}_{eg}~\hat{\sigma}_{g'''g''}\nonumber\\
&&\hat{c}^{k}_{g'e'}[\tilde{\mathcal{H}}_{nh}]^{-1}_{e'e'''}\hat{\mathcal{A}}^{j'+}_{e'''g'''}.
\eea
Note that in writing Eq. (\ref{eq38}) we have interchanged the indices $j,j'$ while defining $\hat{\mathcal{M}}_{g''g'}$ as they are just running indices. 

Now separating $\hat{\mathcal{M}}$ into a Hermitian $(h)$ and anti-Hermitian $(ah)$ part as $\hat{\mathcal{M}}_{gg'} = [\mathcal{M}_{gg'}]_{h}+i[\mathcal{M}_{gg'}]_{ah}$ we get
\bea
\label{eq39}
\dot{\hat{\sigma}}_{g'g} & = &\frac{i}{\hbar}[\mathcal{H}_{c},\hat{\sigma}_{g'g}]-i\sum_{g''}\bigg[\hat{\sigma}_{g'g''}\bigg([\hat{\mathcal{M}}_{g''g}]_{h}\nonumber\\
&-&i[\hat{\mathcal{M}}_{g''g}]_{ah}\bigg)-\bigg([\hat{\mathcal{M}}_{g'g''}]_{h}+i[\hat{\mathcal{M}}_{g'g''}]_{ah}\bigg)\hat{\sigma}_{g''g}\bigg]\nonumber\\
&+&\sum_{k} \sum_{jj'}\sum_{ee'}\sum_{e'' e'''}\sum_{g''g'''}\hat{\mathcal{A}}^{j-}_{g''e''}[\tilde{\mathcal{H}}^\dagger_{nh}]^{-1}_{e''e}\hat{c}^{k\dagger}_{eg}~\hat{\sigma}_{g'''g''}\nonumber\\
&&\hat{c}^{k}_{g'e'}[\tilde{\mathcal{H}}_{nh}]^{-1}_{e'e'''}\hat{\mathcal{A}}^{j'+}_{e'''g'''}.
\eea
Furthermore, writing $[\hat{\mathcal{M}}_{gg'}]_{h} = \frac{1}{2}\left(\hat{\mathcal{M}}_{gg'}+\hat{\mathcal{M}}^{\dagger}_{gg'} \right)$ and $[\hat{\mathcal{M}}_{gg'}]_{ah} = \frac{1}{2}\left(\hat{\mathcal{M}}_{gg'}-\hat{\mathcal{M}}^{\dagger}_{gg'} \right)$ we get 
\begin{widetext}
\bea
\label{eq40}
\dot{\hat{\sigma}}_{g'g} & = &\frac{i}{\hbar}[\mathcal{H}_{c},\hat{\sigma}_{g'g}]-\frac{i}{2}\sum_{g''}\bigg[\hat{\sigma}_{g'g''}\bigg(\hat{\mathcal{M}}_{g''g}+\hat{\mathcal{M}}^{\dagger}_{g''g}\bigg)-\bigg(\hat{\mathcal{M}}_{g'g''}+\hat{\mathcal{M}}^{\dagger}_{g'g''}\bigg)\hat{\sigma}_{g''g}\bigg]+\frac{i}{2}\sum_{g''}\bigg[i\hat{\sigma}_{g'g''}\bigg(\hat{\mathcal{M}}_{g''g}-\hat{\mathcal{M}}^{\dagger}_{g''g}\bigg)\nonumber\\
&&+i\bigg(\hat{\mathcal{M}}_{g'g''}-\hat{\mathcal{M}}^{\dagger}_{g'g''}\bigg)\hat{\sigma}_{g''g}\bigg]+\sum_{k} \sum_{jj'}\sum_{ee'}\sum_{e'' e'''}\sum_{g''g'''}\hat{\mathcal{A}}^{j-}_{g''e''}[\tilde{\mathcal{H}}^\dagger_{nh}]^{-1}_{e''e}\hat{c}^{k\dagger}_{eg}~\hat{\sigma}_{g'''g''}\hat{c}^{k}_{g'e'}[\tilde{\mathcal{H}}_{nh}]^{-1}_{e'e'''}\hat{\mathcal{A}}^{j'+}_{e'''g'''}.
\eea
Now noting that in general one can write $(\hat{\mathcal{M}}\pm\hat{\mathcal{M}}^{\dagger}) = \sum_{jj'}\hat{\mathcal{A}}^{j'-}\left[\tilde{\mathcal{H}}^{-1}_{nh}\pm(\tilde{\mathcal{H}}^{-1}_{nh})^\dagger\right]\hat{\mathcal{A}}^{j+}$ and defining an effective Hamiltonian in the form
\bea
\label{eq41}
\hat{\mathcal{H}}_{eff}= \frac{1}{2}\sum_{jj'}\hat{\mathcal{A}}^{j'-}\left[\tilde{\mathcal{H}}^{-1}_{nh}+(\tilde{\mathcal{H}}^{-1}_{nh})^\dagger\right]\hat{\mathcal{A}}^{j+} +\hat{\mathcal{H}}_{c},
\eea
we can reduce Eq. (\ref{eq40}) after some algebra to the following form
\bea
\label{eq42}
\dot{\hat{\sigma}}_{g'g} & = &\left[\mathcal{H}_{eff}, \hat{\sigma}_{g'g}\right]-\frac{1}{2}\sum_{k}\sum_{jj'}\sum_{ee'}\sum_{e'' e'''}\sum_{g''}\bigg(\hat{\mathcal{A}}^{j'-}_{g'e''}[\tilde{\mathcal{H}}^\dagger_{nh}]^{-1}_{e''e}\left(\hat{c}^{k\dagger}_{eg}\hat{c}^{k}_{g'e'}\right)[\tilde{\mathcal{H}}_{nh}]^{-1}_{e'e'''}\hat{\mathcal{A}}^{j+}_{e''' g''}\hat{\sigma}_{g''g}+\hat{\sigma}_{g'g''}\hat{\mathcal{A}}^{j'-}_{g''e''}[\tilde{\mathcal{H}}^\dagger_{nh}]^{-1}_{e''e}\nonumber\\
&&\left(\hat{c}^{k\dagger}_{eg}\hat{c}^{k}_{g'e'}\right)[\tilde{\mathcal{H}}_{nh}]^{-1}_{e'e'''}\hat{\mathcal{A}}^{j+}_{e''' g}\bigg)+\sum_{k} \sum_{jj'}\sum_{ee'}\sum_{e'' e'''}\sum_{g''g'''}\hat{\mathcal{A}}^{j-}_{g''e''}[\tilde{\mathcal{H}}^\dagger_{nh}]^{-1}_{e''e}\hat{c}^{k\dagger}_{eg}~\hat{\sigma}_{g'''g''}\hat{c}^{k}_{g'e'}[\tilde{\mathcal{H}}_{nh}]^{-1}_{e'e'''}\hat{\mathcal{A}}^{j'+}_{e'''g'''}.
\eea 
\end{widetext}
In writing the above equation we have used the identity $\hat{\mathcal{A}}^{j'-}_{ge}\left[\mathcal{H}^{-1}_{nh}-(\mathcal{H}^{-1}_{nh})^\dagger\right]_{ee'}\hat{\mathcal{A}}^{j+}_{e'g'} = \hat{\mathcal{A}}^{j'-}_{ge}\sum_{e''e'''}(\mathcal{H}^{-1}_{nh})^\dagger_{ee''}\left[\hat{c}^{k\dagger}_{e''g''}\hat{c}^{k}_{g'''e'''}\right](\hat{\mathcal{H}}^{-1}_{nh})_{ee'}\hat{\mathcal{A}}^{j+}_{e'g'}$ \cite{Reiter12}. 

If we now define an effective Lindblad operator, that represents decay from one ground-state to another in the subspace $M_g$, in the form 
\bea
\label{eq43a}
\mathcal{L}^{k}_{eff} = \sum_{j'}\sum_{g'}\sum_{ee'}\hat{c}^{k}_{ge}[\hat{\mathcal{H}}^{-1}_{nh}]_{ee'}\hat{\mathcal{A}}^{j'+}_{e'g'}, 
\eea
we can reduce Eq. (\ref{eq42}) to that of a Liouvillian master equation. In doing so attention must be towards the ordering of the operators. So far we have been careful about retaining normal ordering for all the involved photon annihilation and creation operators such that positive frequency operators always appear to the left of negative frequency operators, c.f. the discussion following Eq. (7). This ordering ensures e.g., that there is no dynamics when all incident fields are in vacuum. For evaluating the final expressions this ordering should be kept. As a consequence of this ordering, however, the last term in Eq. (\ref{eq42}) does not have the form corresponding to standard matrix multiplication where repeated indices are summed, since we cannot rearrange non-commuting terms. To write Eq. (\ref{eq42}) in a more appealing form we rearrange the terms but introduce the normal ordering $:...:$ to indicate that $\mathcal{E}^+$ and $\mathcal{E}^-$ should always be evaluated in the normal ordered form. Hence, keeping this in mind, Eq. (\ref{eq42}) on using Eq. (\ref{eq43a}) takes the form of a standard Liouvillian master equation:
\bea
\label{eq43b}
\dot{\hat{\sigma}} & = &:i\left[\mathcal{H}_{eff}, \hat{\sigma}\right]-\frac{1}{2}\sum_{k}\bigg(\mathcal{L}^{k\dagger}_{eff}\mathcal{L}^{k}_{eff}\sigma+\sigma\mathcal{L}^{k\dagger}_{eff}\mathcal{L}^{k}_{eff}\bigg)\nonumber\\
&+&\sum_{k}\mathcal{L}^{k}_{eff}\hat{\sigma}\mathcal{L}^{k\dagger}_{eff}~:~.
\eea 
The operator $\sigma$ here represents the population and coherences involving the ground states of the emitters only. 

The master Eq. (\ref{eq43b}) gives an effective equation for the ground state coherences. The solution to this equation can be directly substituted into Eq. (\ref{eq25}) in order to describe the full  evolution, e.g., in a system where the ground-state population evolves due to Raman scattering from one  state to another. This result is a direct generalization of the EO equation obtained in Ref. \cite{Reiter12} to the Heisenberg picture and to quantum fields. Here the main new feature appearing is the normal ordering, which gives the correct prescription for how to treat quantum fields. For classical fields in a coherent state, the normal ordering can be removed and the result reduces to that of Ref. \cite{Reiter12}. In the following paper \cite{Dasp2}, we give some examples of how to use this result together with Eq. (\ref{eq25}) to describe the scattering dynamics. 
\section{Summary}
In summary, we have developed a general Heisenberg picture formalism to study photon scattering from a system of multi-level emitters embedded in a $3$-dimensional dielectric medium. Our formalism directly gives the output field in terms of the input field and the system's ground-state dynamics. To find the ground-state evolution, we have derived an effective operator master equation in the Heisenberg picture Eq. (\ref{eq43b}). Together these two expressions allow a full description of the evolution of the system and is a generalization of the EO approach of Ref. \cite{Reiter12} to photon scattering. 

The key assumption in this work is that the incident intensity of the fields is sufficiently low that we can ignore saturation effect of the emitters. In this approximation we directly obtain the scattering relation in Eq. (\ref{eq25}), which describe the scattering of individual photons, but does not contain the direct (fast) optical nonlinearity associated with multiple photons incident on the emitters at the same time. Through Eq. (\ref{eq43b}) the formalism does, however, accommodate the (slow) optical nonlinearity between photons incident at different times  e.g. through optical pumping of the ground states. The restriction on the validity is thus only in the incoming intensity and not in the total incident number of photons. 

We emphasize the compactness of our formalism and its ability to provide an exact solution of the reflected and transmitted amplitudes of the scattered photon. In principle, our formalism can solve for any complex intra- and inter-emitter dynamics, provided that the non-Hermitian Hamiltonian can be inverted. Being completely general our formalism can be applicable to a plethora of systems where emitters are coupled to a dielectric medium including $1$D waveguides. We show this explicitly in part II of this series \cite{Dasp2} where we consider several different kinds of emitter configurations coupled to a double-sided $1$D waveguide.
\begin{acknowledgments}
This work was support by the ERC Grant QIOS (Grant No. 306576) and the Danish Council for Independent Research (Natural Science). 
FR gratefully acknowledges financial support from the Humboldt Foundation. 
\end{acknowledgments}


\end{document}